\documentclass[prl,preprint,showpacs,amssymb,superscriptaddress]{revtex4}
\usepackage{graphicx}% Include figure files
\usepackage{dcolumn}% Align table columns on decimal point
\usepackage{bm}% bold math
\begin{document}

\date{\today}

\title{Further Evidence for the Decay $K^+ \to \pi^+ \nu \bar\nu$}

\author{S.~Adler} \affiliation{Brookhaven National Laboratory, Upton,
New York 11973} \author{A.O.~Bazarko} \affiliation{Joseph Henry
Laboratories, Princeton University, Princeton, New Jersey 08544}
\author{P.C.~Bergbusch} \affiliation{Department of Physics and
Astronomy, University of British Columbia, Vancouver, British
Columbia, Canada, V6T 1Z1} \author{E.W.~Blackmore}
\affiliation{TRIUMF, 4004 Wesbrook Mall, Vancouver, British Columbia,
Canada, V6T 2A3} \author{D.A.~Bryman} \affiliation{Department of
Physics and Astronomy, University of British Columbia, Vancouver,
British Columbia, Canada, V6T 1Z1} \author{S.~Chen}
\affiliation{TRIUMF, 4004 Wesbrook Mall, Vancouver, British Columbia,
Canada, V6T 2A3} \author{I-H.~Chiang} \affiliation{Brookhaven National
Laboratory, Upton, New York 11973} \author{M.V.~Diwan}
\affiliation{Brookhaven National Laboratory, Upton, New York 11973}
\author{J.S.~Frank} \affiliation{Brookhaven National Laboratory,
Upton, New York 11973} \author{J.S.~Haggerty} \affiliation{Brookhaven
National Laboratory, Upton, New York 11973} \author{J.~Hu}
\affiliation{TRIUMF, 4004 Wesbrook Mall, Vancouver, British Columbia,
Canada, V6T 2A3} \author{T.~Inagaki} \affiliation{High Energy
Accelerator Research Organization (KEK), Oho, Tsukuba, Ibaraki
305-0801, Japan} \author{M.~Ito}\altaffiliation{Present address:
Thomas Jefferson
National Accelerator Facility, Newport News, Virginia 23606.}
\affiliation{Joseph Henry Laboratories,
Princeton University, Princeton, New Jersey 08544}
\author{V.~Jain}
\affiliation{Brookhaven National Laboratory, Upton, New York 11973}
\author{S.~Kabe} \affiliation{High Energy Accelerator Research
Organization (KEK), Oho, Tsukuba, Ibaraki 305-0801, Japan}
\author{S.H.~Kettell} \affiliation{Brookhaven National Laboratory,
Upton, New York 11973} \author{P.~Kitching} \affiliation{Centre for
Subatomic Research, University of Alberta, Edmonton, Canada, T6G 2N5}
\author{M.~Kobayashi} \affiliation{High Energy Accelerator Research
Organization (KEK), Oho, Tsukuba, Ibaraki 305-0801, Japan}
\author{T.K.~Komatsubara} \affiliation{High Energy Accelerator
Research Organization (KEK), Oho, Tsukuba, Ibaraki 305-0801, Japan}
\author{A.~Konaka} \affiliation{TRIUMF, 4004 Wesbrook Mall, Vancouver,
British Columbia, Canada, V6T 2A3} \author{Y.~Kuno}
\altaffiliation{Present address: Department of Physics, Osaka
University, Toyonaka, Osaka 560-0043, Japan.}  \affiliation{High
Energy Accelerator Research Organization (KEK), Oho, Tsukuba, Ibaraki
305-0801, Japan} \author{M.~Kuriki} \affiliation{High Energy
Accelerator Research Organization (KEK), Oho, Tsukuba, Ibaraki
305-0801, Japan} \author{K.K.~Li} \affiliation{Brookhaven National Laboratory,
Upton, New York 11973}
\author{L.S.~Littenberg} \affiliation{Brookhaven National Laboratory,
Upton, New York 11973} \author{J.A.~Macdonald} \affiliation{TRIUMF,
4004 Wesbrook Mall, Vancouver, British Columbia, Canada, V6T 2A3}
\author{ P.D.~Meyers} \affiliation{Joseph Henry Laboratories,
Princeton University, Princeton, New Jersey 08544}
\author{J.~Mildenberger} \affiliation{TRIUMF, 4004 Wesbrook Mall,
Vancouver, British Columbia, Canada, V6T 2A3} \author{M. Miyajima}
\affiliation{Department of Applied Physics, Fukui University, 3-9-1
Bunkyo, Fukui, Fukui 910-8507, Japan} \author{N.~Muramatsu}
\altaffiliation{Present address: Japan Atomic Energy Research
Institute, Sayo, Hyogo 679-5198, Japan.}
\affiliation{High Energy Accelerator Research Organization (KEK), Oho,
Tsukuba, Ibaraki 305-0801, Japan} \author{T.~Nakano}
\affiliation{Research Center for Nuclear Physics, Osaka University,
10-1 Mihogaoka, Ibaraki, Osaka 567-0047, Japan} \author{C.~Ng}
\altaffiliation{Also at Physics Department, State University of New
York at Stony Brook, Stony Brook, NY 11794-3800.}
\affiliation{Brookhaven National Laboratory, Upton, New York 11973}
\author{S.~Ng} \affiliation{Centre for
Subatomic Research, University of Alberta, Edmonton, Canada, T6G 2N5}
\author{T.~Numao} \affiliation{TRIUMF, 4004 Wesbrook Mall, Vancouver,
British Columbia, Canada, V6T 2A3} \author{J.-M.~Poutissou}
\affiliation{TRIUMF, 4004 Wesbrook Mall, Vancouver, British Columbia,
Canada, V6T 2A3} \author{R.~Poutissou} \affiliation{TRIUMF, 4004
Wesbrook Mall, Vancouver, British Columbia, Canada, V6T 2A3}
\author{G.~Redlinger} \altaffiliation{Present address: Brookhaven National Laboratory.}
\affiliation{TRIUMF, 4004 Wesbrook Mall, Vancouver, British Columbia,
Canada, V6T 2A3} \author{T.~Sato} \affiliation{High Energy Accelerator
Research Organization (KEK), Oho, Tsukuba, Ibaraki 305-0801, Japan}
\author{K.~Shimada} \affiliation{Department of Applied Physics, Fukui
University, 3-9-1 Bunkyo, Fukui, Fukui 910-8507, Japan}
\author{T.~Shimoyama} \affiliation{Department of Applied Physics,
Fukui University, 3-9-1 Bunkyo, Fukui, Fukui 910-8507, Japan}
\author{T.~Shinkawa} \altaffiliation{Present address: National Defense
Academy of Japan, Yokosuka, Kanagawa 239-8686, Japan.}
\affiliation{High Energy Accelerator Research Organization (KEK), Oho,
Tsukuba, Ibaraki 305-0801, Japan} \author{F.C.~Shoemaker}
\affiliation{Joseph Henry Laboratories, Princeton University,
Princeton, New Jersey 08544} \author{J.R.~Stone} \affiliation{Joseph
Henry Laboratories, Princeton University, Princeton, New Jersey 08544}
\author{R.C.~Strand} \affiliation{Brookhaven National Laboratory,
Upton, New York 11973} \author{S.~Sugimoto} \affiliation{High Energy
Accelerator Research Organization (KEK), Oho, Tsukuba, Ibaraki
305-0801, Japan} \author{Y.~Tamagawa} \affiliation{Department of
Applied Physics, Fukui University, 3-9-1 Bunkyo, Fukui, Fukui
910-8507, Japan} \author{C.~Witzig} \affiliation{Brookhaven National
Laboratory, Upton, New York 11973} \author{Y.~Yoshimura}
\affiliation{High Energy Accelerator Research Organization (KEK), Oho,
Tsukuba, Ibaraki 305-0801, Japan} \collaboration{E787 Collaboration}

\newpage
\begin{abstract}
Additional evidence for the rare kaon decay $K^+ \to \pi^+ \nu
\bar\nu$ ~ has been found in a new data set with comparable
sensitivity to the previously reported result.  One new event was
observed in the pion momentum region examined, $211<P<229$ MeV/$c$,
bringing the total for the combined data set to two.  Including all
data taken, the backgrounds were estimated to contribute $0.15\pm0.05$
events.  The branching ratio is $B$($K^+ \to \pi^+ \nu
\bar\nu$)$=1.57^{+1.75}_{-0.82} \times 10^{-10}$.
\end{abstract}

\maketitle

The decay $K^+ \to \pi^+ \nu \bar\nu$~ is very sensitive to the
coupling of top to down quarks, $V_{td}$, in the
Cabibbo-Kobayashi-Maskawa  quark mixing matrix. In the context of
the Standard Model (SM), the predicted branching ratio is $B(K^+ \to \pi^+
\nu \bar\nu)=0.75 \pm 0.29\times 10^{-10}$ \cite{newbb}.  In an
earlier study, a single event consistent with the decay $K^+ \to \pi^+
\nu \bar\nu$ ~ at a branching ratio of $B$($K^+ \to \pi^+ \nu
\bar\nu$)= $1.5^{+3.5}_{-1.2}\times10^{-10}$ was found\cite{pnn99,pnn95}.
In this letter, final results from Experiment E787\cite{det,E949} at the
Alternating Gradient Synchrotron (AGS) of Brookhaven National
Laboratory are presented, including a new data sample of comparable
sensitivity to that reported previously.

The observable signature for $K^+ \! \rightarrow \! \pi^+ \nu
\overline{\nu}$ decay from kaons at rest involves only the $\pi^+$
track and $\pi^+$ decay products.  Major background sources include
the two-body decays $K^+ \!  \rightarrow \! \mu^+ \nu_\mu$ ($K_{\mu
2}$) and $K^+ \!  \rightarrow \!  \pi^+ \pi^0$ ($K_{\pi 2}$),
pions scattered from  the beam, and $K^+$ charge exchange (CEX) reactions
resulting in decays $K_L^0\to\pi^+ l^- \overline{\nu}_l$, where $l=e$
or $\mu$. In order to make an unambiguous measurement of $B(K^+ \to
\pi^+ \nu \bar\nu)$, it is advantageous to suppress all backgrounds
well below the signal level.

The data discussed here were acquired during the 1998 run of the AGS,
using kaons of 710~MeV/$c$ incident on the apparatus at a rate of
about $4$ MHz.  The kaons were detected and identified by
\v{C}erenkov, tracking, and energy-loss counters after which 27\%
reached a scintillating-fiber target used for kaon and pion tracking.
Measurements of the momentum ($P$), range ($R$),
and kinetic energy ($E$) of charged decay products were
made using the target, a central drift chamber, and a cylindrical
range stack (RS) made up of 21 layers of plastic scintillator with two
layers of tracking chambers embedded in it, all within a 1-T
solenoidal magnetic field.  The $\pi^+ \!  \rightarrow \! \mu^+ \!
\rightarrow \!  e^+$ decay sequence from pions which came to rest in
the RS was observed using 500-MHz transient digitizers.  Photons were
detected in a calorimeter mainly consisting of a
14-radiation-length-thick barrel detector made of lead/scintillator
sandwich and 13.5-radiation-length-thick endcaps of undoped CsI
crystals.

To be accepted as a $K^+ \!  \rightarrow \! \pi^+ \nu \overline{\nu}$
candidate, a decay particle must be positively identified as a $\pi^+$
by comparing $P$, $R$, and $E$ measurements, and by observation of the $\pi^+
\!  \rightarrow \! \mu^+ \!  \rightarrow \!  e^+$ decay sequence.
Events containing other decay products including photons or beam
particles were eliminated by detectors covering $4\pi$ sr.  A clean
hit pattern in the scintillating-fiber target and a delayed decay at
least 2~ns after an identified $K^+$ suppressed background events due
to CEX and scattered beam pions.  The search was restricted to the
measured momentum region $211< P <229$ MeV/$c$ between the
$K_{\mu 2}$ and $K_{\pi 2}$ peaks. The maximum pion momentum from $K^+
\!  \rightarrow \! \pi^+ \nu \overline{\nu}$ decays at rest is 227
MeV/$c$.

The data analysis, described in Refs.\cite{pnn99,Paul}, focused on
obtaining detailed estimates of all backgrounds prior to examining the
pre-determined signal region.  In order to evaluate observed events,
the parameter space of observables for the 1998 data set was
subdivided into 7500 bins with differing levels of expected
backgrounds. The signal region was defined to include the first 486
bins and was not examined until the final step in the analysis
procedure. The background expected in the signal region was estimated
from $BG=\Sigma^{486}_{i=1} b_i$ where $b_i$ is the expected number of
background events from all sources in bin $i$. Assuming the SM value
for the branching ratio, $7.5 \times 10^{-11}$\cite{sm_assump}, an
expected signal number $S_i$ was also obtained for each bin as $S_i=
7.5 \times 10^{-11} A_i N_K$ where $A_i$ is the acceptance in bin $i$
and $N_K$ is the number of kaons.  A signal-to-background function
$f=\frac{S_i}{b_i} $ was defined to characterize each bin in terms of
the relative probability for events occurring there to originate in
$K^+ \!  \rightarrow \! \pi^+ \nu \overline{\nu}$ decay or
background. For the event observed in the 1995-97 data set (Event A)
which passed the tightest cuts designed to evaluate candidate events,
a similar procedure resulted in the signal-to-background function
value $f^A=35$, which indicated a very high probability that it was
due to signal and a low level of consistency with any of the known
sources of background.

For the 1998 data set, the final candidate selection requirements were
similar to those used previously, although more stringent track
reconstruction criteria were imposed.  For the background sources
listed above, the numbers of events expected were $n_{{K_{\mu 2}}} =
0.034^{+0.043}_{-0.024}$, $n_{{K_{\pi 2}}} = 0.012^{+0.003}_{-0.004}$,
$n_{Beam} = 0.004\pm 0.001$ and $n_{CEX} = 0.016^{+0.005}_{-0.004}$,
where the combined statistical and systematic uncertainties are
given. In total, the background level anticipated in the signal region
was $BG=0.066^{+0.044}_{-0.025}$ events.  For this level of
background, the acceptance for $K^+ \to \pi^+ \nu \bar\nu$ was
$A=\Sigma^{486}_{i=1} A_i = 0.00196 \pm 0.00005^{stat} \pm
0.00010^{syst}$, obtained from the factors given in the last column of
Table~\ref{acceptance}.  The estimated systematic uncertainty in the
acceptance was due mostly to the uncertainty in pion-nucleus
interactions.

\begin{table}
\begin{tabular}{|l|l|l|}
\hline
~Acceptance factors & 1995--97& 1998 \\
\hline
$K^+$ stop efficiency &   $0.704 $& $0.702$ \\ 
$K^+$ decay after 2 ns &  $0.850 $& $0.851$ \\ 
$K^+ \to \pi^+ \nu \bar\nu$~ phase space &   $ 0.155 $& $0.136$ \\ 
Solid angle acceptance &   $0.407 $& $0.409$ \\ 
$\pi^+$ stop efficiency & $0.513$& $0.527$ \\ 
Reconstruction efficiency &          $0.959 $& $0.969$\\ 
Other kinematic constraints &              $0.665 $& $0.554$ \\ 
$\pi - \mu - e$ decay acceptance &     $ 0.306 $& $0.392 $\\ 
Beam and target analysis &           $0.699 $& $0.706 $\\ 
Accidental loss &         $0.785 $& $0.751 $\\ 
\hline 
Total acceptance &   $0.0021 $& $0.00196$ \\ 
\hline 
\end{tabular}
\caption{\label{acceptance}Acceptance factors used in the
measurement of $K^+ \to \pi^+ \nu \bar\nu$ for the 1995-7 and 1998
data sets.  The ``$K^+$ stop efficiency'' is the fraction of kaons
entering the target that stopped, and 
``$\pi^+$
stop efficiency'' is the fraction of pions which stopped  in the
range stack without nuclear interactions or decay-in-flight.
``Other kinematic constraints''
includes  particle identification cuts. }
\end{table}

To confirm the background estimates, the selection criteria were
relaxed\cite{pnn95,Paul} to allow about 14 times higher background,
and all bins except those in the final signal region were
examined. Two events were observed, in agreement with the number of
expected background events $\Sigma^{7500}_{i=487} b_i=0.9\pm 0.7$ for
this region.  One of these (Event B1) had a low value of the
signal-to-background function $f^{B1}=0.07$.  The second event
(Event B2) had all the characteristics of a signal event
but a low apparent time of $\pi\rightarrow\mu$ decay resulted in an
intermediate value for the signal-to-background function $f^{B2}=0.7$
reflecting the possibility that it was due to $K_{\mu 2}$ decay\cite{pme}.

\begin{figure}
\includegraphics*[width=\linewidth]{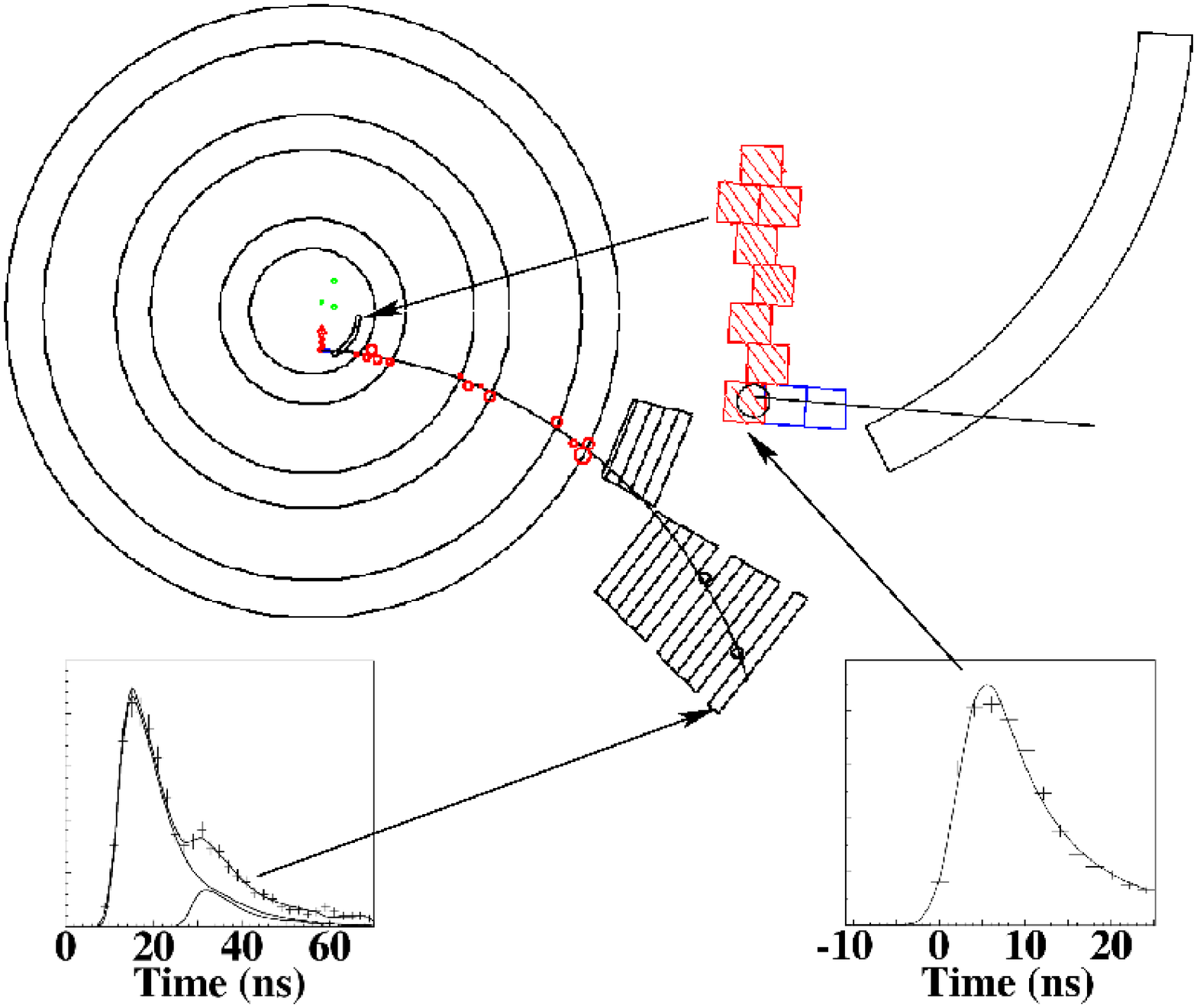}
%{/nunubar/home/joseph/pass2/98/paper/evtdsp_xxx.eps}
\caption{ Display of candidate Event C.  On the top left is the end
view of the detector showing the track in the target, the drift
chamber, and the range stack. On the top right is a blow-up
of the track in the target,
where the  hatched squares represent
target fibers hit by the $K^+$ and the open squares indicate those hit by
the $\pi^+$; a trigger scintillator that was hit is also shown.
The lower right hand box shows the
digitized signal in the target fiber where the kaon stopped indicating
no additional activity. The pulse
was sampled every 2 ns (crosses) and the solid line is a fit.  The lower
left hand box shows the digitized $\pi\rightarrow\mu$  decay signal in
the scintillator where the pion stopped. The curves are fits for the
first, second and combined pulses.}
\label{EventC}
\end{figure}

Following the background study, the signal region for the 1998 data
set was examined yielding one candidate event (Event C) with
$f^{C}=3.6$. The kinematic values are $P=213.8 \pm 2.7$~MeV/$c$,
$R=33.9 \pm 1.2$~cm (in equivalent cm of
scintillator), and $E=117.1 \pm 3.6$~MeV. A display of
this event is shown in Fig.~\ref{EventC}. Close inspection of Event C
indicates that it is consistent with being due to $K^+ \to \pi^+ \nu
\bar\nu$ decay.  Although a small amount (0.6 MeV in total) of
coincident nontrack related energy was observed in the target, this
level of additional energy was expected to occur randomly somewhere in
the detector in about 11\% of all events, and was well below the cut
at 3 MeV on this parameter.

\begin{figure}
\includegraphics*[width=\linewidth]{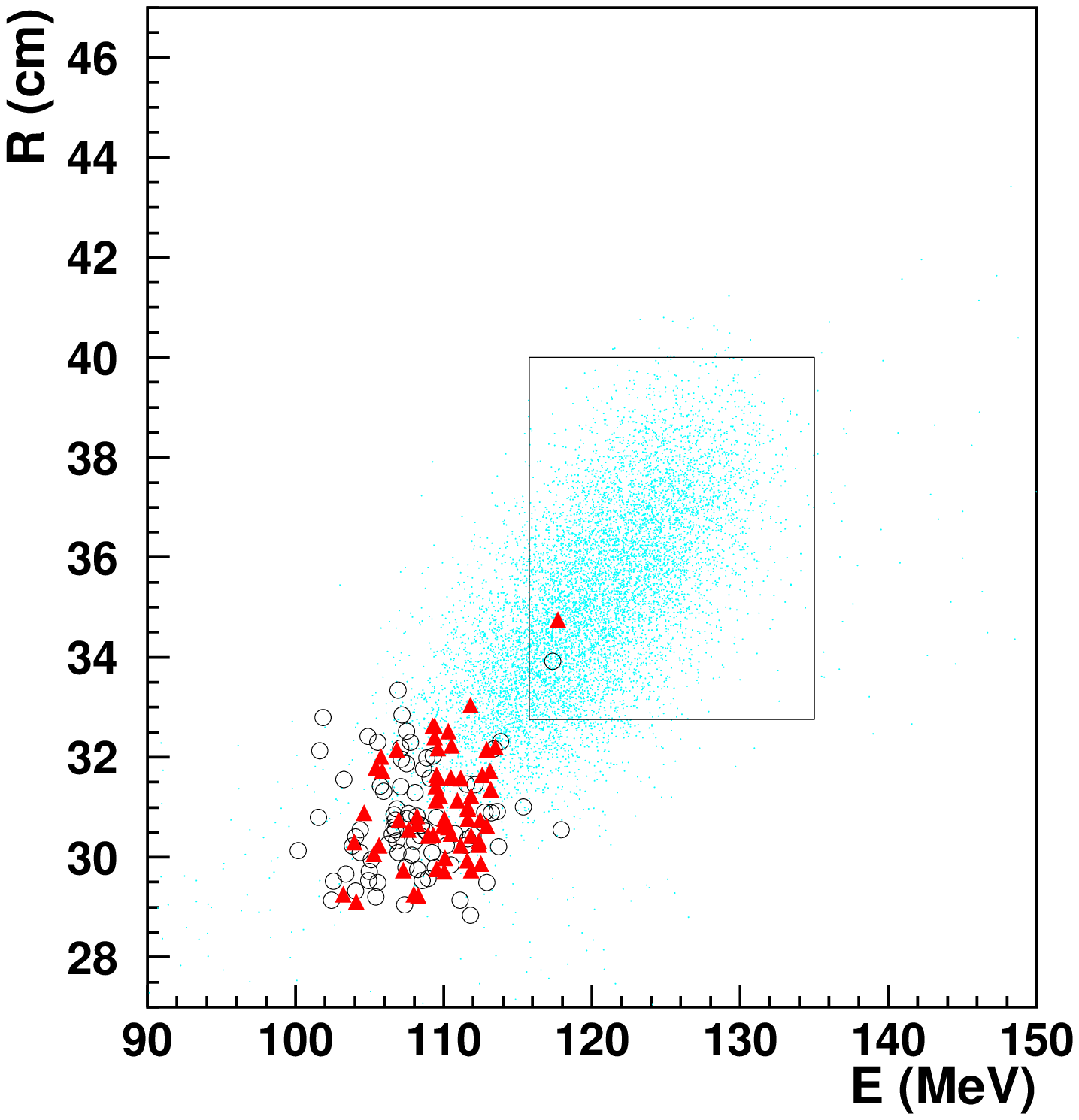}
%{box_new_fullmc.epsi}
\caption{ Range  vs. energy plot of the
final sample. The circles are for the 1998 data and the triangles are
for the 1995-97 data set. The group of events around $E=108$~MeV is
due to the $K_{\pi 2}$ background.  The simulated distribution of
expected events from $K^+ \!  \rightarrow \!  \pi^+ \nu
\overline{\nu}$ is indicated by dots.}
\label{rve}
\end{figure}

The combined result for E787 data taken between 1995 and 1998 is
shown in Fig.~\ref{rve}, the range vs. kinetic energy of events
surviving all other cuts. In Fig.~\ref{rve} the box represents the
signal region in which two events (Events A and C) appear.  Using the
$f$ values of the observed events, a likelihood ratio
technique\cite{Junk} was used to determine the best estimate of the
branching ratio.  Based on two observed events with their associated
$f$ values, the acceptances given in Table ~\ref{acceptance}, the
numbers of $K^+$ incident on the target, and the expected background
levels given in Table~\ref{Nbg}, the result is $B(K^+ \!  \rightarrow
\!  \pi^+ \nu \overline{\nu}) = 1.57^{+1.75}_{-0.82} \times
10^{-10}$\cite{L90,outside_box}.  This result would be consistent with
being due entirely to background only at the level of
0.02\%\cite{Junk}.

\begin{table}
\begin{tabular}{|l|l|l|}
\hline
~& 1995--97& 1998 \\
\hline
&& \\
$N_{K} $  &$3.2 \times10^{12}$ & $2.7\times 10^{12}$\\
Observation (events)& 1 & 1  \\
Estimated background (events)& $0.08\pm 0.03$ & $0.066^{+0.044}_{-0.025}$\\
\hline 
\end{tabular}
\caption{\label{Nbg} Numbers of kaons incident on the target,
observed events, and
total estimated background events expected  
for the 1995-97 and 1998 data samples.}
\end{table}

Bounds on $|V_{td}|$ may be obtained from $B(K^+ \to \pi^+
\nu\bar\nu)$ by maximizing the charm quark contribution within the
limits given in Ref.\cite{newbb} and assuming its phase relative to
the top quark contribution to be $0^\circ$ or $180^\circ$.  The 
branching ratio limits given above along with $\bar m_t(m_t) = 166
\pm 5\,$GeV/$c^2$, and $V_{cb} =0.041\pm0.002$\cite{newbb} yield $0.007
<|V_{td}| <0.030$
(68\% C. L.)\cite{V90}.
Note that these limits do not require
knowledge of $V_{ub}$ or $\epsilon_K$.  Alternatively, one can extract
corresponding limits
on the quantity $|\lambda_t|$ ($\lambda_t \equiv
V^*_{ts}V_{td}$): $2.9 \times 10^{-4} < |\lambda_t| <1.2 \times
10^{-3}$.  In addition, the  bounds $-0.88 \times 10^{-3} <
Re(\lambda_t) < 1.2 \times 10^{-3}$  can be obtained.
For $Im(\lambda_t)$, an upper limit of $Im(\lambda_t) < 1.1 \times
10^{-3}$
(90\% C.L)
is found.
The bounds on $\lambda_t$ are derived
without reference to the $B$ system or to measurements of $\epsilon_K$
or $\epsilon'/\epsilon$ and are of particular interest because
$Im(\lambda_t)$ is proportional to the area of the unitarity triangle.

The limit found in the search for decays of the form $K^+ \!
\rightarrow \!  \pi^+ X^0$, where $X^0$ is a neutral weakly
interacting massless particle \cite{x0}, is $B(K^+ \!  \rightarrow \!
\pi^+ X^0) < 0.59 \times 10^{-10}$ (90\% CL), based on zero events
observed in a $\pm 2 \sigma$ region around the pion kinematic
endpoint.

We would like to acknowledge the contributions made by colleagues who
participated in earlier phases of this work including M. Atiya,
T. F. Kycia (deceased), D. Marlow, and A. J. S. Smith.  We gratefully
acknowledge the dedicated effort of the technical staff supporting
this experiment and of the Brookhaven AGS Department.  This research
was supported in part by the U.S. Department of Energy under Contracts
No. DE-AC02-98CH10886, W-7405-ENG-36, and grant DE-FG02-91ER40671, by
the Ministry of Education, Culture, Sports, Science and Technology of
Japan through the Japan-U.S. Cooperative Research Program in High
Energy Physics and under the Grant-in-Aids for Scientific Research for
Encouragement of Young Scientists and for JSPS Fellows, and by the
Natural Sciences and Engineering Research Council and the National
Research Council of Canada.


\begin{thebibliography}{ (a) }
\bibitem{E949} An upgraded version of the experiment, E949, is
pursuing the study of $K^+ \to \pi^+ \nu
\bar\nu$.
\bibitem{sm_assump} The value assumed here is arbitrary and
does not affect the result obtained for the branching ratio.
\bibitem{pme} Fluctuations in the shapes of single RS counter pulses
as recorded by the transient digitizers can occasionally produce false
$\pi \rightarrow \mu$-like signals appearing to come at early decay times.
\end{thebibliography}
\end{document}